\documentclass[draftcls,onecolumn,a4paper,12pt]{IEEEtran}

\usepackage{amsmath}
\usepackage{amssymb}
\usepackage{bm}
\usepackage{graphicx}
\usepackage{epstopdf}
\usepackage{mathtools}
\usepackage{algorithm}
\usepackage{algorithmic}
\usepackage{caption}
\usepackage{subcaption}

\newcommand{\me}{\ensuremath{\mathrm{e}}}
\newcommand{\PLi}{\ensuremath{\text{PL}_{\text{dB},i}}}
\DeclarePairedDelimiter\ceil{\lceil}{\rceil}
\DeclarePairedDelimiter\floor{\lfloor}{\rfloor}

\begin{document}

\title{Subcarrier-Chunk Assignment With Power Allocation and
  Multiple-Rate Constraints for Downlink OFDMA}

\author{Juthatip Wisanmongkol and Wiroonsak Santipach,~\IEEEmembership{Senior~Member,~IEEE}%
\thanks{The material in this paper was presented in part at the ISCIT, Republic of Korea, 2014~\cite{iscit14}.}%
\thanks{J. Wisanmongkol is currently with National Electronics and Computer
  Technology Center (NECTEC), 112 Phahonyothin Road, Khlong Nueng,
  Khlong Luang District, Pathumthani 12120, Thailand.  J. Wisanmongkol was previously with and W. Santipach is with the Department of Electrical Engineering;
  Faculty of Engineering; Kasetsart University, Bangkok, 10900,
  Thailand (email: wiroonsak.s@ku.ac.th).}}

\maketitle

\begin{abstract}
We consider downlink OFDMA in which users are imposed with different
proportional-rate constraints, and a base station would like to
maximise user rate over subcarrier assignment (SA) and power
allocation (PA). To attain the objective, we propose a chunk-based SA
and low-complexity PA algorithms.  Numerical results show that the
proposed SA and PA can provide higher average minimum achievable rate
than the existing schemes while keeping the deviation from the rate
constraint small for both single-cell and multi-cell systems with
fractional frequency reuse.
\end{abstract}

\section{Introduction}

In orthogonal frequency division multiple access (OFDMA), mobile users
can be dynamically assigned to transmit in different sets of
subcarriers based on their channel qualities.  An approach called rate
adaptive allocation is commonly used to dynamically allocate resource
to maximise a total throughput.  To achieve this goal, only the user
with the best channel quality is assigned subcarriers for data
transmission.  However, this results in an unfair assignment to other
users.

Fair resource allocation was proposed in various work concerning both
wireless and wireline channels.  In~\cite{conf:rhee}, Rhee and Cioffi
proposed an algorithm that achieves proportional-rate fairness through
maximising the minimum user's throughput.  However, this approach is
inapplicable to a system in which users have different requested
rates.  To address this issue, Shen {\em et al.}~\cite{journal:shen}
and Ren {\em et al.}~\cite{journal:ren} formulated a nonlinear
optimisation problem that guarantees multiple-rate fairness.
However, determining the solution to this nonlinear problem can be
computationally complex.  One way to reduce complexity is to assign
subcarriers to users in a group or chunk of adjacent subcarriers in
OFDMA networks~\cite{journal:zhu, journal:zhu2}.  Since the number of
channel-filter taps is much lower than the number of total
subcarriers, adjacent subcarriers are highly correlated.  Therefore,
if the subcarriers are appropriately grouped together and selected for
users, the resulting achievable rate can approach that of a
single-subcarrier-based resource allocation at a lower computational
cost.

For this work (and our conference proceeding~\cite{iscit14}), we
propose a chunk-based subcarrier assignment (SA) based
on~\cite{conf:rhee} and~\cite{journal:shen}.  For a given subcarrier
assignment, we propose a power allocation (PA) based on a low
signal-to-noise ratio (SNR) approximation.  With this approximation,
the problem is linearized, and its solution can be easily obtained.
The chunk-based allocation problem we consider here is significantly
different from that in~\cite{journal:zhu, journal:zhu2} which consider
average BER (bit error rate) constraint instead of proportional-rate
constraint.  Since there is no equivalent chunk-based SA scheme
existing in the literature, we modify the single-subcarrier based
scheme by~\cite{journal:shen} to assign subcarriers in chunk for the
purpose of performance comparison.  Numerical results show that our
proposed allocation scheme improve the average minimum user's
achievable rate over the modified scheme based on~\cite{journal:shen}
in all SNR regimes.  The rate gain is larger when SNR is low.

We extend our work in~\cite{iscit14} to multi-cell channels in which
intercell interference can be significant.  To mitigate the
interference and improve cell-edge performance, intercell interference
coordination (ICIC) such as frequency reuse planning is adopted to
restrict resource allocation among different cells in the
network~\cite{journal:a.s.hamza}.  By using a frequency reuse factor
of 1 (FRF-1), high peak data rate can be achieved at a cost of high
interference levels at the cell edges.  On the contrary, by using a
frequency reuse factor of more than 1 (or less than 1, depending on
the notation), interference level can be reduced at a cost of a lower
spectrum utilisation~\cite{book:t.s.rappaport}.  Fractional frequency
reuse (FFR) is a promising ICIC technique for OFDMA network wherein
cells are partitioned into regions with different
FRF's~\cite{boudreau09} and is proposed for next-generation wireless
systems~\cite{techreport:ffrintro}.  We apply the proposed chunk-based
SA with uniform PA in multi-cell setting with FFR and show that the
resulting rate is larger than static SA by as much as 400\%.

\section{Single-Cell Resource Allocation}
\label{sec:su}

First we consider a single-cell downlink OFDMA network, which consists
of a base station and $K$ mobile users.  This model is applicable to a
cellular network with a large frequency-reuse factor, in which
intercell interference is negligible.  The channel between the base
station and any user $k$ is assumed to be frequency-selective fading
with order $\ell_k$ for which channel-filter taps are denoted by
$\{h_{k,0}, h_{k,1}, ..., h_{k,\ell_k-1}\}$.  Assuming $N$ total
subcarriers, the frequency response of user $k$ on subcarrier $n$ can
be obtained by taking a discrete Fourier transform (DFT) of the
channel impulse response as follows
\begin{equation}
  H_{k,n} = \sum_{i=0}^{\ell_k - 1} h_{k,i} \, \me^{\frac{-j 2 \pi i
      n}{N}} .
\label{hnk}
\end{equation}
For data transmission from the base station, each user is assigned
chunk(s) of contiguous subcarriers.  For 3GPP LTE-Advanced, the
minimum subcarrier assignment unit is one resource block, which
consists of 12 subcarriers.  Each chunk of subcarriers is exclusively
used by a single user; therefore, there is no interference from other
users in the cell.  Let $M$ be the total number of subcarrier chunks,
which is greater than or equal to the number of users ($M \geq K$).
Hence, each user will be assigned at least one chunk.  The number of
chunks $M = \floor*{\frac{N}{L}}$ where $L$ is a chunk size and
remaining subcarriers will be appended to the last chunk.

The sum achievable rate for user $k$ is given by
\begin{equation}
  R_k = \sum_{m=1}^M \omega_{k,m} R_{k,m}
\end{equation}
where $\omega_{k,m} \in \{0, 1\}$ indicates whether user $k$ is
assigned to transmit in chunk $m$.  Since only one user is assigned to
transmit in each subcarrier chunk at any moment, for $1 \le m \le M$,
\begin{equation}
  \sum_{k=1}^K \omega_{k,m} = 1.
\end{equation}
Because the data signal in each subcarrier is only corrupted by
additive white Gaussian noise with zero mean and variance
$\sigma^2_w$, the achievable rate per subcarrier for user $k$ in chunk
$m$ per subcarrier is given by
\begin{equation}
  R_{k,m} = \frac{1}{N} \sum_{n = (m-1)L+1}^{m L} \log_2 \left(1 +
  \frac{p_{k,n} |H_{k,n}|^2}{\sigma^2_w} \right)
\end{equation}
where $p_{k,n}$ is the transmission power allocated for user $k$ in
subcarrier $n$.

The objective of the proposed resource allocation is to maximise the
sum achievable rate of all users over chunk assignment and power
allocation, subject to total-power and proportional-rate constraints.
Assuming that users can have different rate requirement, we let
$\gamma_1 : \gamma_2 : \cdots : \gamma_K$ be a ratio of the requested
rates of user 1 through user $K$.  Thus, the optimisation problem can
be stated as follows:

\begin{align}
\max_{\{\omega_{k,m}\},\{p_{k,n}\}} &\sum_{k=1}^{K} R_{k} \label{eq:objective}\\
\text{subject to} & \sum_{k=1}^{K} \sum_{n=1}^{N} p_{k,n} \leq P_T, \\
& p_{k,n} \geq 0, \quad \text{for } 1 \leq k \leq K \text{ and } 1 \leq n
\leq N, \\
& \sum_{k=1}^{K} \omega_{k,m} = 1, \quad \text{for } 1 \leq m \leq M, \\
& \omega_{k,m} \in \{0, 1\}, \\
& R_{1}:R_{2}:\ldots:R_{K} = \gamma_{1}:\gamma_{2}:\ldots:\gamma_{K} . 
\label{eq_fair}
\end{align}

Finding the solution to this integer-nonlinear problem is
prohibitively complex.  Also, due to a strict fairness constraint
in~\eqref{eq_fair}, solutions may not exist at all.  To reduce
complexity of the problem, we propose to find a suboptimal solution by
breaking the problem into two subproblems.  The first subproblem is SA
in which uniform transmit power over all subcarriers is assumed.
Then, with the specific sets of subcarrier assignment obtained from
solving the first subproblem, PA is subsequently performed.  From the
numerical results, when the number of chunks is sufficiently large, a
mere subcarrier assignment assuming uniform power can roughly satisfy
the proportional-rate constraints

\subsection{Subcarrier Assignment}
\label{sec:sa}

For both SA's proposed by~\cite{conf:rhee} and~\cite{journal:shen}, in
the first iteration, subcarriers are sequentially assigned to the
users based on channel gains.  Thus, the subcarrier with the highest
gain is assigned to the first user.  This gives undue favour to the
first user and as a result, the first user generally has the highest
average rate whereas the last user generally has the lowest average
rate.  This discrepancy becomes more apparent with larger chunk size.
Moreover, subcarrier assignment based only on absolute channel gain
might not result in the largest sum throughput.  To reduce the
mentioned discrepancy, our proposed SA removes the serial assignment
in the first iteration, and changes the assignment criterion from the
achievable rate of a user to the normalised rate defined as a ratio
between rate of a user and average rate of all users for that chunk.
The normalised rate of user $k$ in chunk $m$ is given by
\begin{equation}
  \bar{R}_{k,m} = \frac{R_{k,m}}{\frac{1}{K} \sum_{k'=1}^{K} R_{k',m}}
  .
\end{equation}
For the given subcarrier assignment, uniform power allocation across
all subcarriers is assumed, i.e., $p_{k,n} = \frac{P_T}{N}$.

In the first iteration of the proposed algorithm, the base station
registers the chunks with the highest normalised rate for each user
$k$.  To maximise the minimum user's rate, the registered chunk with
the smallest normalised rate over its requested rate will be selected
and assigned to the corresponding user.  The user that has already
been assigned a chunk will be removed from the assignment pool $U$.
This step is repeated until all users are assigned one chunk each.
The algorithm is stated in Algorithm~\ref{al_sub}.  In subsequent
iterations, all users are back in the assignment pool.  In each
iteration, the user with the smallest $R_k/\gamma_k$ is assigned the
chunk that maximises the normalised rate over all the remaining
chunks.  This step is iterated until all chunks are assigned.  We
denote the set of chunks assigned to user $k$ by $\Theta_k$.  Thus,
Algorithm~\ref{al_sub} gives nonoverlapping sets $\Theta_1, \Theta_2,
\cdots, \Theta_K$ of which their union spans all chunks.  Let $N_k$ be
the number of subcarriers assigned to user $k$.  Therefore,
$\sum_{k=1}^K N_k = N$.

\begin{algorithm}
  \caption{Subcarrier assignment (SA)}
  \label{al_sub}
  \begin{algorithmic}[1]
		
    \STATE Set $S = \{1,2,\ldots, M\}$ and $U = \{1, 2, \ldots, K\}$
    where $M \ge K$.

    \STATE Set $\Theta_{k} = \emptyset, \forall k \in U$.		

    \STATE Set $R_k = 0, \forall k \in U$.		

    \STATE Find $\bar{R}_{k,m}, \forall k \in U \text{ and } m \in S$ .	

    \WHILE{$U \neq \emptyset$}		
	\FOR{$k \in U$}
	\STATE Find $$ m^*_k = \arg \max_{m \in S} \bar{R}_{k,m} .$$
	\ENDFOR

    \STATE Find $$ k^* = \arg \min_{k \in U}
    \frac{\bar{R}_{k,m^*_k}}{\gamma_k} .$$

    \STATE Update $\Theta_{k^*} \gets \Theta_{k^*} \cup \{m^*_{k^*}\}$
    and $S \gets S \setminus \{m^*_{k*}\}$.
		
    \STATE Update $R_{k^*} \gets R_{k^*} + R_{k^*,m^*_{k^*}}$ and $U
    \gets U \setminus \{k^*\}$.
		
    \ENDWHILE
		
    \STATE Reset $U = \{1,2,\ldots, K\}$.
		
    \WHILE{$S \neq \emptyset$}
		
    \STATE Find $$k' = \arg \min_{k \in U} \, \frac{R_{k}}{\gamma_{k}}
    .$$		

    \STATE Find $$m'_{k'} = \arg \max_{m \in S} \, \bar{R}_{k',m} .$$
		
    \STATE Update $\Theta_{k'} \gets \Theta_{k'} \cup \{m'_{k'} \}$
    and $S \gets S \setminus \{m'_{k'} \}$.
		
    \STATE  Update $R_{k'} \gets R_{k'} + R_{k',m'_{k'}}$.
    \ENDWHILE
\end{algorithmic}
\end{algorithm}

To determine the complexity of the proposed SA algorithm and compare
with the algorithm proposed by~\cite{journal:shen}, we count the
number of $\log$-computations and the number of comparisons required
in the algorithm.  We note that finding the extremum of $n$ entries
requires at most $n$ comparisons.  Both the proposed SA algorithm and
the algorithm by~\cite{journal:shen} needs to know the instantaneous
rate of each subcarrier for each user and thus, requires $KN$
$\log$-computations.

For the proposed SA, there are $\sum_{i=1}^K i(M-K+i)$ comparisons in
the initial phase and $\sum_{i=1}^{M-K} M + 1 -i$ comparisons in the
second phase.  Thus, the total number of comparisons increases as
$\mathcal{O}(MK^2)$ as $M$ and $K$ increase.  For the
single-subcarrier-based SA by~\cite{journal:shen}, the initial phase
requires $\sum_{i = 1}^K N - K +i$ while the second phase requires
$\sum_{i=1}^{N-K} K + i$ comparisons.  Thus, the number of comparisons
increases as $\mathcal{O}(N^2)$.  For a moderate and large chunk size,
$N$ could be much greater than $M$ and $K$ and the algorithm
in~\cite{journal:shen} will require much larger number of comparisons
than our chunk-based algorithm.

\subsection{Power Allocation}

Given subcarrier allocation obtained by Algorithm~\ref{al_sub}, the
problem in~\eqref{eq:objective} is reduced to

\begin{align}
\max_{\{p_{k,n}\}} &\sum_{k=1}^{K} R_k \label{eq:reduced} \\
\text{subject to} & \sum_{k=1}^{K} \sum_{n \in \Omega_k} p_{k,n} \leq P_T \\
& p_{k,n} \geq 0, \quad \text{for } 1 \leq k \leq K \text{ and } 1 \leq n
\leq N \\
& R_{1}:R_{2}:\ldots:R_{K} = \gamma_{1}:\gamma_{2}:\ldots:\gamma_{K} .
\end{align}

This PA for each subcarrier was solved by~\cite{journal:shen}.  Since
we will borrow some definitions from~\cite{journal:shen}, the solution
by~\cite{journal:shen} will be briefly described first.  For the
subcarrier assignment of user $k$ denoted by $\Omega_k$, each
subcarrier is ordered by a ratio of its squared channel magnitude to
the noise power $G_{k,(n)} \triangleq |H_{k,(n)}|^2 / \sigma^2_w$ in
an increasing order, i.e., $G_{k,(1)} \le G_{k,(2)} \le \cdots \le
G_{k,(N_k)}$.  The optimal power for user $k$ is then computed
by~\cite{journal:shen}
\begin{equation}
p_{k,(n)} = p_{k,(1)} + \frac{G_{k,(n)} -
	G_{k,(1)}}{G_{k,(n)}G_{k,(1)}}
\label{eq:powersubcarrier}
\end{equation}
and the total power allocated for user $k$ is given by
\begin{equation}
  P_{T,k} = \sum_{n=1}^{N_{k}} p_{k,(n)} = N_{k}p_{k,(1)} + V_k
\label{eq:powertotal}
\end{equation}
where
\begin{equation}
  V_{k} = \sum_{n=2}^{N_{k}}
  \frac{G_{k,(n)}-G_{k,(1)}}{G_{k,(n)}G_{k,(1)}} .
\end{equation}
To find the set of optimal total power allocated to all users $\{
P_{T,k}\}$, the following nonlinear
system needs to be solved~\cite{journal:shen}
\begin{multline}
  \frac{N_{1}}{\gamma_1} \left\{\log_2 \left(1 +
  G_{1,(1)}\frac{P_{T,1}-V_{1}}{N_{1}}\right)+\log_2 W_{1}\right\}
  \\ = \frac{N_{k}}{\gamma_k} \left\{\log_2 \left(1 +
  G_{k,(1)}\frac{P_{T,k}-V_{k}}{N_{k}}\right)+\log_2 W_{k}\right\},
  \forall k
\label{eq:poweropti}
\end{multline}
and
\begin{equation}
  \sum_{k=1}^{K} P_{T,k} = P_{T}
\label{eq:powercstr}
\end{equation}
where for $2 \le k \le K$,
\begin{equation}
  W_{k} =
  \left(\prod_{n=2}^{N_{k}}\frac{G_{k,(n)}}{G_{k,(1)}}\right)^\frac{1}{N_{k}}
  .
\end{equation}
Solving~\eqref{eq:poweropti} and~\eqref{eq:powercstr} can be complex
and requires some numerical methods.  To simplify the PA problem, we
propose to linearize the nonlinear system and thus, reduce the
complexity of the problem.  We apply a low-SNR approximation and
obtain a suboptimal but linear PA problem.  This low-SNR approximation
is well justified since power allotted for each subcarrier is usually
small due to a large number of subcarriers.  Per subsequent numerical
examples, the proposed solution also performs well even for a
moderate-SNR system.

The ratio between the actual and desired rates can be approximated as
follows
\begin{align}
\frac{R_k}{\gamma_k} &= \frac{1}{\gamma_k N} \sum_{n_k \in \Omega_k}
\log_2(1 + \frac{1}{\sigma^2_w} p_{k, n_k} |H_{k,n_k}|^2)\\
&\approx \frac{\log_2(\me)}{\gamma_k \sigma^2_w N} \sum_{n_k \in
	\Omega_k} p_{k, n_k} |H_{k,n_k}|^2 .
\label{eq:approx}
\end{align}
For~\eqref{eq:approx}, we assume low-SNR regime, i.e., $P_T/\sigma^2_w
\ll 1$, and apply the approximation $\log_2(1+x) \approx x
\log_2(\me)$ when $x \ll 1$.  With~\eqref{eq:approx} and the
proportional rate constraint
\begin{equation}
  \frac{R_1}{\gamma_1} = \frac{R_2}{\gamma_2} = \cdots = \frac{R_K}{\gamma_K},
\end{equation}
we obtain a linear system with $K$ equations and $K$ unknowns with the
following matrix equation
\begin{equation}
\label{eq:powermatrix}
  \begin{bmatrix}
    1 & 1 & 1 & \cdots & 1 \\ 
    1 & \alpha_{2} & 0 & \cdots & 0 \\ 
    1 & 0 & \alpha_3 & \cdots & 0 \\ 
    \vdots & \vdots & \vdots & \ddots & \vdots \\ 
    1 & 0 & 0 & \cdots & \alpha_{K} \\
\end{bmatrix}
\left[
\begin{array}{c}
  P_{T,1} \\ P_{T,2} \\ P_{T,3} \\ \vdots \\ P_{T,K}
\end{array}
\right]
=
\left[
\begin{array}{c}
  P_{T} \\ \beta_{2} \\ \beta_3 \\ \vdots \\ \beta_{K}
\end{array}
\right]
\end{equation}
where 
\begin{align}
  \alpha_{k} &= -\frac{\gamma_{1} E_{k} N_{1}G_{k,(1)}}{\gamma_{k}
    E_{1} N_{k} G_{1,(1)}},\\ 
   E_{k} &= \sum_{n=1}^{N_{k}} \frac{G_{k,(n)}}{G_{k,(1)}},
\end{align}
and
\begin{multline}
  \beta_{k} = \frac{\gamma_{1}E_{k}N_{1}}{\gamma_{k}E_{1}G_{1,(1)}} -
  \frac{\gamma_{1}N_{1}N_{k}}{\gamma_{k}E_{1}G_{1,(1)}} -
  \frac{\gamma_{1}E_{k}N_{1}G_{k,(1)}}{\gamma_{k}E_{1}N_{k}G_{1,(1)}}
  V_{k} \\
 + \frac{N_{1}}{G_{1,(1)}} \left(\frac{N_{1}}{E_{1}}-1 \right)
  + V_{1}.
\end{multline}

The solution of the linear system~\eqref{eq:powermatrix} can be easily
obtained as follows
\begin{align}
  P_{T,1} &= \frac{P_T - \sum_{k = 2}^K \beta_k/\alpha_k}{1 - \sum_{k =
      2}^K 1/\alpha_k}, \label{eq_pt1}\\ 
  P_{T,k} &= \frac{1}{\alpha_k}(\beta_k -
  P_{T,1}), \quad \forall k \ne 1 . \label{eq_ptk}
\end{align}

We remark that the solution presented arises from the low-SNR
approximation and sometimes may not be feasible, i.e., some powers are
negative.  To remedy this negative-power solution, we propose to
allocate uniform power for the group of users with the smallest powers
(including all users with negative power).  The number of users in
this group will be just large enough that their combined transmit
power exceeds zero.  All other users not in this group will be
allocated power according to the solution in~\eqref{eq_pt1}
and~\eqref{eq_ptk}.  The steps of the proposed power allocation are
shown in Algorithm~\ref{alg_pw}.  If $P_{T,k} > 0$ for all $k$, the
proposed PA is straightforward. 
\begin{algorithm}
  \caption{Power allocation (PA)}
  \label{alg_pw}
  \begin{algorithmic}[1]
    \STATE For each subcarrier assignment $\Omega_k$, obtain $\{
    G_{k,(n)} \}$.
		
    \STATE Determine $\alpha_k$ and $\beta_k$ for $2 \le k \le K$.
		
    \STATE Solve~\eqref{eq_pt1} and~\eqref{eq_ptk} to obtain $\{P_{T,
      k}\}$.
		
    \IF{$\exists P_{T, k} < 0$}
	
        \STATE Arrange $\{P_{T,k}\}$ in ascending order: $P_{T,(1)}
        \le P_{T,(2)} \le \cdots \le P_{T,(K)}$.
        
        \STATE Set $k = 1$.

        \STATE Set $P_{\text{sum}} = P_{T,(k)}$.

        \WHILE{$P_{\text{sum}} < 0$}
           \STATE Update $k \gets k + 1$.
           
           \STATE Update $P_{\text{sum}} \gets P_{\text{sum}} +
           P_{T,(k)}$.  
        \ENDWHILE

        \FOR{$i \in \{1, 2,\ldots, k\}$}
             \STATE Update $P_{T,(i)} \gets \frac{P_{\text{sum}}}{k}$.
        \ENDFOR

     \ENDIF
		
     \WHILE{$\exists P_{T, k} < V_{k}$}
		
     \STATE Update $\Omega_k \gets \Omega_k \setminus \{\arg \min_{n_k
       \in \Omega_k} G_{k,(n_k)} \} $ and $N_k \gets N_k - 1$.
		
     \STATE Update $V_k$.
		
     \ENDWHILE
		
     \STATE For each $\Omega_k$, compute $p_{k,(n)}$
     from~\eqref{eq:powersubcarrier} and~\eqref{eq:powertotal}.
  \end{algorithmic}
\end{algorithm}
We note that if $P_{T, k} < V_{k}$, then according
to~\eqref{eq:powertotal}, $p_{k,n} < 0$.  To find a feasible solution,
the subcarriers with smallest channel gains will be allocated zero
power.  Thus, subcarriers whose channel gains are below some threshold
will not be use for transmission, similar to a water-filling scheme.
This process is repeated until $P_{T,k} \geq V_k$.

Reference~\cite{journal:shen,conf:i.c.wong} also proposed to
linearize~\eqref{eq:poweropti} but with different approach.  Both work
assume that the ratio between the number of subcarriers assigned to
each user and its requested rate is fixed for all users,
$\frac{N_k}{\gamma_k} = \text{const.}, \forall k$.  This assumption
does not often hold true and hence, may not be as practical as the
low-SNR assumption.  The resulting linear system appeared
in~\cite{journal:shen,conf:i.c.wong} differs
from~\eqref{eq:powermatrix}, but has similar structure.  When SNR is
high, reference~\cite{journal:shen,conf:i.c.wong} made another
approximation for~\eqref{eq:poweropti}, but still ended up with
nonlinear equation.

\section{Multi-Cell Subcarrier Assignment}

\begin{figure}
  \centering
  \begin{subfigure}[b]{0.25\linewidth}
    \includegraphics[width=1in]{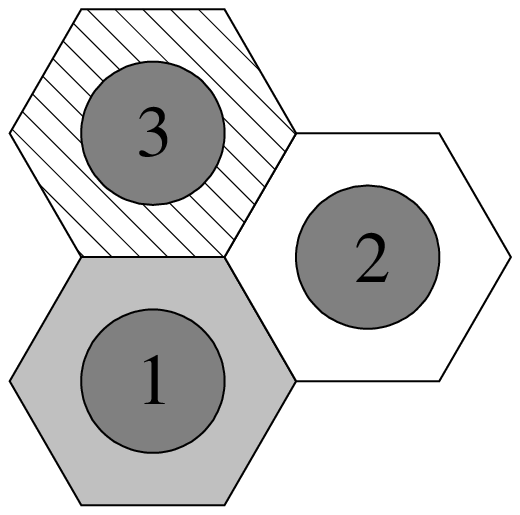}
    \label{fig:ffr1}
  \end{subfigure}
  \quad
  \begin{subfigure}[t]{0.7\linewidth}
    \includegraphics[width=2.3in]{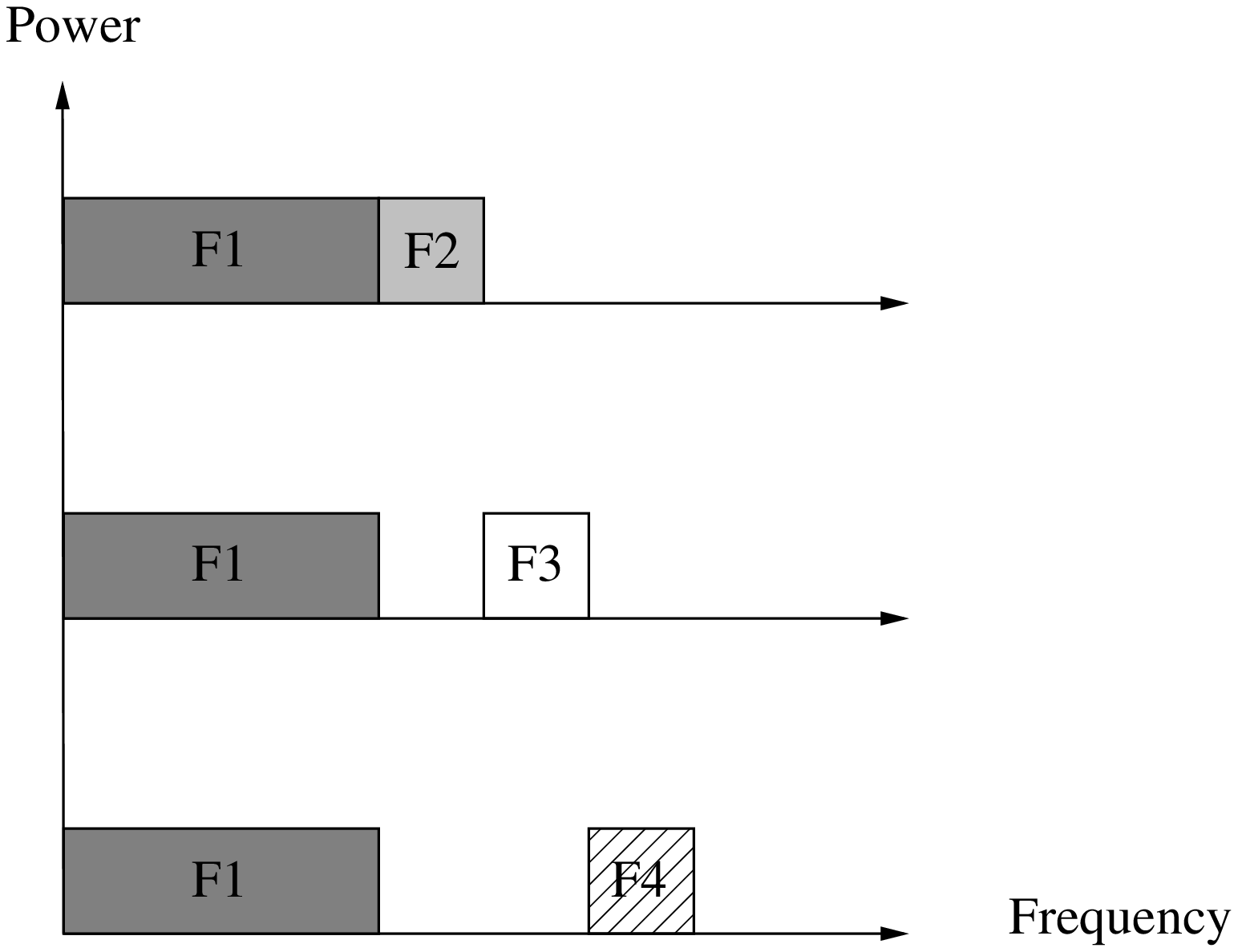}
    \label{fig:ffr2}
  \end{subfigure}
  \caption{Fractional frequency reuse with $\text{FRF} =
    3$.}\label{fig:ffrall}
\end{figure}

For a multi-cell channel, FFR is applied as follows.  First, available
transmission bandwidth is divided into two bands, namely cell-centre
and cell-edge bands, as shown in Fig~\ref{fig:ffrall}.  Since users in
a cell-centre region is less susceptible to intercell interference,
the frequency reuse factor for the cell-centre group is set to 1 to
enhance spectrum efficiency.  The bandwidth range for cell-centre
users is denoted by F1.  However, the reuse factor for cell-edge users
is set to be 3 with 3 different nonoverlapping frequency ranges
denoted by F2, F3, and F4.

Group membership can be determined by a threshold based on the
received signal strength or the distance between the user and the base
station denoted by $d_u$~\cite{conf:h.lei}.  Assuming that the cell
radius is $R$ and the cell-centre radius is $\tau$, users with
distance $d_u \leq \tau$ away from the base station will be in a
cell-centre group while users with distance $d_u > \tau$ will be in a
cell-edge group.  Finding a proper threshold or radius of the cell
centre, $\tau$, is usually heuristic~\cite{conf:t.novlan}.  Assuming
that there are $K$ users in each cell, we denote the number of
cell-centre users and cell-edge users by $K_{cc}$ and $K_{ce}$,
respectively, and $K_{cc} + K_{ce} = K$.  Let us denote the set of
users in the cell-centre group by $\Omega_{cc}$ and that in cell-edge
group by $\Omega_{ce}$.  For a network with uniformly distributed
users, the optimal number of subcarriers allocated to the cell-centre
group $N_{cc}$ and the cell-edge group $N_{ce}$ is proportional to a
coverage area~\cite{letter:z.bharucha} and is given by
\begin{align}
  N_{cc} &= \ceil*{N \left(\frac{\tau}{R}\right)^2}, \\ 
  N_{ce} &= \floor*{\frac{N-N_{cc}}{\text{FRF}}} .
\end{align}
Assuming that chunk size is fixed at $L$, the number of chunks for
cell-centre and cell-edge users are given by $M_{cc} =
\floor*{\frac{N_{cc}}{L}}$ and $M_{ce} = \floor*{\frac{N_{ce}}{L}}$,
respectively.

\begin{figure}[h]
  \centering 
  \includegraphics[width=3in]{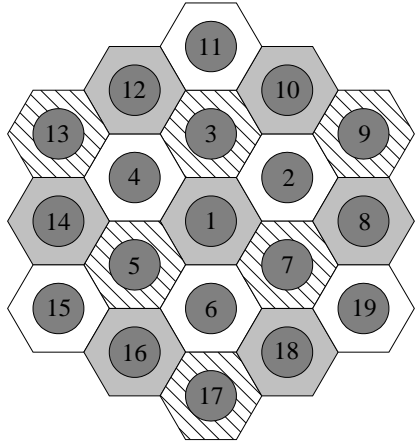}
  \caption{A two-tier 19-cell model.}
  \label{fig:cellular}
\end{figure}

We consider a 2-tier 19-cell network shown in Fig.~\ref{fig:cellular}.
With the proposed FFR, a user in cell-centre area is interfered by all
18 other cells while a user in cell-edge area is interfered by 6 other
cells only.
We approximate the SINR for user $k$ in the cell-centre area of cell 1
on the $n$th subcarrier as follows
\begin{equation}
  \text{SINR}_{k,n} \approx \frac{|H_{k,n,1}|^2 P_T/N}{\sigma^2_n +
    \sum_{i=2}^{19} 10^{-0.1 \text{PL}_{\text{dB},i}} |H_{k,n,i}|^2
    P_T/N}, \quad k \in \Omega_{cc}
\end{equation}
where $|H_{k,n,i}|^2$ is a squared channel gain from base station $i$
and can be computed from channel impulse response similar
to~\eqref{hnk} and propagation path loss in decibel for interfering
signal from base station $i$ is given by~\cite{conf:h.lei}
\begin{equation}
  \PLi = 128.1 + 37.6 \log_{10}(R_i)
\end{equation}
where $R_i$ is the distance in kilometre from base station $i$ to base
station 1.  The approximation discards path loss of the desired signal
due to the distance from base station 1 to the user since that
distance is much shorter than $R_i$.  We also assume uniform power
allocation over all subcarriers since adaptive power allocation for
all cells in the network is not practical due to required complex
coordination among base stations.  Moreover,
references~\cite{journal:jang} and \cite{journal:g.song} suggest that
improvement from adaptive power allocation is marginal over a wide
range of SNR's when only subcarriers with high gain are selected.
Similarly, the SINR of cell-edge user $k$ in cell 1 on subcarrier $n$
is approximated by
\begin{multline}
  \text{SINR}_{k,n}\approx \\
  \frac{|H_{k,n,1}|^2 P_T/N}{\sigma^2_n +
    \sum_{\substack{i \in \{8, 10, 12,\\ 14, 16, 18\}}} 10^{-0.1
      \text{PL}_{\text{dB},i}} |H_{k,n,i}|^2 P_T/N}, \quad k \in
  \Omega_{ce} .
\end{multline}
Given target BER, an effective sum rate per subcarrier in chunk $m$
for user $k$ can be computed by
\begin{equation} 
   R_{k,m} = \frac{1}{N} \sum_{n=(m-1)L+1}^{mL} \log_{2} \left( 1 +
   \lambda \text{SINR}_{k,n} \right)
\end{equation}
where $\lambda = -1.5/\ln(5
\text{BER})$~\cite{conf:h.lei,techreport:macrocell} and the sum rate
for user $k$ in either cell-centre or cell-edge areas over all chunks
is given by
\begin{equation}
  R_{k} = \left\{ \begin{array}{l@{\quad:\quad}l} \sum_{m=1}^{M_{cc}}
    \omega_{cc,k,m} R_{k,m}& k \in \Omega_{cc}\\ \sum_{m=1}^{M_{ce}}
    \omega_{ce,k,m} R_{k,m}& k \in \Omega_{ce}
 \end{array} \right.
\end{equation}
where indication functions for user $k$ in chunk $m$ in cell-centre
and cell-edge area are denoted by $\omega_{cc,k,m}$ and
$\omega_{ce,k,m} \in \{0, 1\}$, respectively.  

We would like to maximise the sum throughput for cell 1 while
maintaining proportional-rate fairness among the users in the cell.
Since the transmit power from all base stations is fixed, we assign
chunk of subcarriers to users in cell 1 to maximise the throughput.
The two groups of users may adhere to different proportional-rate
fairness.  The SA problem can be stated as follows
\begin{align}
  \max_{\substack{\{\omega_{cc,k,n}\}\\\{\omega_{ce,k,n} \}}} &\sum_{k \in
    \Omega_{cc} \cup \Omega_{ce}} R_k \label{mpm}\\ 
  \text{subject to} &\sum_{k\in \Omega_{cc}} \omega_{cc,k,m} = 1, \quad \text{for } 1
  \leq m \leq M_{cc}, \\ 
  &\sum_{k \in \Omega_{ce}} \omega_{ce,k,m} = 1, \quad \text{for }
  1 \leq m \leq M_{ce}, \\ 
  &\omega_{cc,k,m}, \omega_{ce,k,m} \in \{0, 1\},\\
   &R_{i_1}:R_{i_2}:\ldots:R_{i_{K_{cc}}} = \gamma_1 : \gamma_2: \ldots : 
     \gamma_{K_{cc}},\\
   &R_{j_1}:R_{j_2}:\ldots:R_{j_{K_{ce}}} = \beta_1: \beta_2: \ldots: 
     \beta_{K_{ce}}.
\end{align}
where $\Omega_{cc} = \{i_1, i_2, \ldots, i_{K_{cc}}\}$, $\Omega_{ce} =
\{j_1, j_2, \ldots, j_{K_{ce}}\}$, and $\gamma_i$ and $\beta_j$ are
the requested rates for cell-centre user $i$ and cell-edge user $j$,
respectively.

The problem stated in~\eqref{mpm} can be divided into two subproblems
and each subproblem maximises the sum rate of users in each group.
Thus, we can apply the proposed SA stated in Algorithm~\ref{al_sub} in
Section~\ref{sec:sa} to solve each subproblem.  The complexity of this
SA follows the discussion at the end of Section~\ref{sec:su} and
increases with the number of users in the cell considered.

\section{Numerical Results}

In this section, the performance of the proposed SA and PA is shown
and compared with existing schemes.  Two main performance indices are
the averaged minimum user's achievable rate and the average rate
constraint deviation, $\bar{\mathcal{D}}$, which indicates how well
the proposed scheme conform to the proportional-rate constraint.
$\bar{\mathcal{D}}$ is defined in~\cite{journal:shen} as
\begin{equation}
\label{eq:ratedeviation}
\bar{\mathcal{D}} = \frac{E\left[\sum_{k=1}^{K} \left|
	\frac{R_{k}}{\sum_{k=1}^{K} R_{k}} -
	\frac{\gamma_{k}}{\sum_{k=1}^{K} \gamma_{k}} \right| \right]}{2 -
	2\min_{1 \le k \le K} \frac{\gamma_{k}}{\sum_{k=1}^{K} \gamma_{k}}}
\end{equation}
where the expectation is over channel realisation.  For the optimal
solution, $\bar{\mathcal{D}} = 0$.

\begin{figure}[h]
  \centering \includegraphics[width=3.72in]{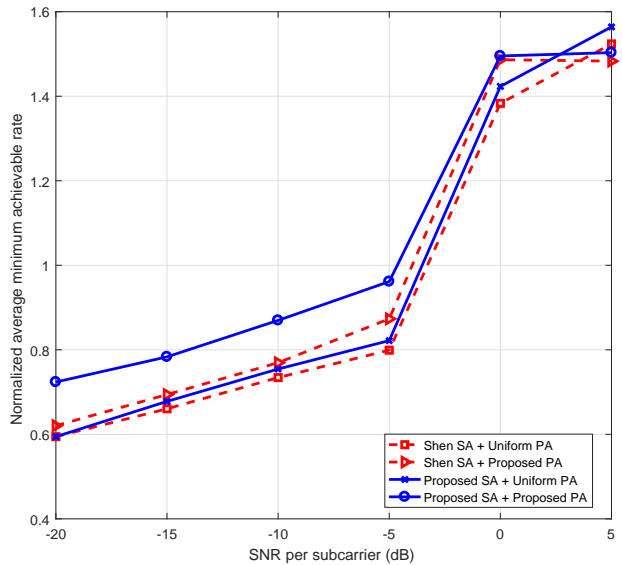}
  \caption{Normalised minimum user's achievable rate is shown with
    different SA and PA schemes with chunk size $L = 1$.  $N=128$,
    $K=4$, and $\gamma_{1}:\gamma_{2}:\gamma_{3}:\gamma_{4} =
    1:1:4:4$.}
  \label{fig:jointrate}
\end{figure}

Our proposed SA scheme can also be applied with chunk size equal to
one ($L = 1$) and thus, can be compared with the SA scheme proposed
by~\cite{journal:shen}.  We assume 4 users in the single-cell system,
which have the requested rates of $\gamma_1:\gamma_2:\gamma_3:\gamma_4
= 1:1:4:4$, and experience Rayleigh fading channels with 4, 8, 16 and
32 taps, respectively.  To compare with the performance of the optimal
allocation, we normalise sum rate of all schemes with the optimal sum
rate.  For reasonable simulation time for the optimal solution, a
system with the number of subcarriers $N=128$\footnote{Subsequent
  examples are shown with much larger $N$.}  is considered.
Fig.~\ref{fig:jointrate} shows a normalised minimum user's achievable
rate with different SA and PA schemes.

The proposed SA offers higher minimum achievable rate when compared
with the scheme by Shen {\em et al.}~\cite{journal:shen} regardless of
PA.  As expected, in a low-SNR region ($\text{SNR} < 0$ dB), the rate
is higher when proposed PA is used since the proposed PA scheme is
derived from a low-SNR approximation.  However, in higher SNR region
($\text{SNR} \ge 0$ dB), the rate is higher when uniform PA is used
instead.  As seen from these results, the rates obtained can be larger
than that of the optimum solution (the normalised rate larger than 1)
since the solution of the proposed scheme might deviate from the rate
constraint. However, we will see in the next figure that the average
deviation is not large.

\begin{figure}[h]
  \centering
  \includegraphics[width=3.72in]{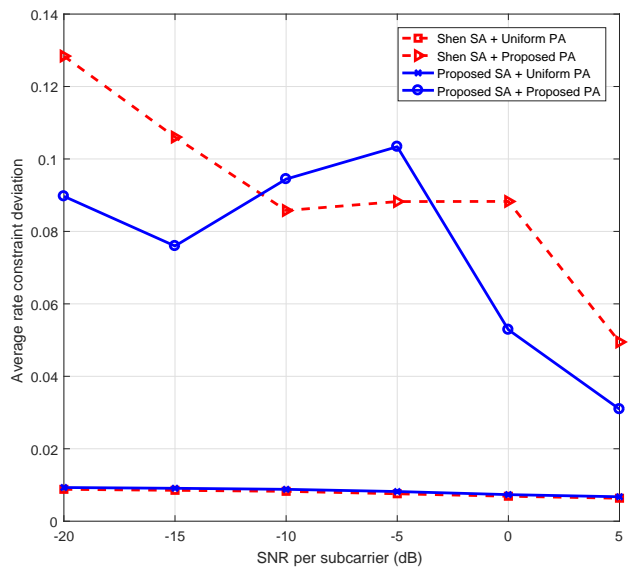}
  \caption{Average rate constraint deviation associated with the rate
    results in Fig.~\ref{fig:jointrate} is shown.}
  \label{fig:jointdev}
\end{figure}

Besides the minimum rate, we also examine how well the rate-fairness
constraint is adhered to.  Fig.~\ref{fig:jointdev} shows the average
rate-constraint deviation associated with the results in
Fig.~\ref{fig:jointrate}.  In a low-SNR regime, the proposed PA has a
relatively higher rate deviation since many subcarriers are assigned
zero power.  This leaves fewer subcarriers for transmission and hence,
proportional-rate constraint is harder to satisfy.  For higher-SNR
regime, rate deviation is lower for all schemes.  We also note that
uniform PA gives lower rate deviation, but with lower sum rate.  Thus,
for a single-subcarrier-based assignment, our proposed SA combined
with uniform PA performs generally well with lesser complexity.

It has been shown in~\cite{conf:hassan} that, for users with the same
target rates, the power consumed is minimal when the frequency reuse
factor in the cell-edge area is 3. Therefore, for the multi-cell
model, we assume a two-tier 19 cells with FRF-1 in the cell-centre
area, and FRF-3 in the cell-edge area as shown in
Fig.~\ref{fig:cellular}.  We assume a total of 8 users per cell,
uniformly distributed within a cell with 1-km radius.  The propagation
channel between a base station and a user follows a 3GPP TR 25.814
macro-cell system~\cite{techreport:macrocell} with the parameters
listed in Table~\ref{tblsimparam}.

\begin{table}
  \caption{Simulation parameters}
  \centering
  \begin{tabular}{|l|c|} \hline
    Parameters & Values \\ \hline
    Path loss & $128.1 + 37.6\log_{10} ( d \text{(km)})$\\
    FFT size & $512$ \\ 
    Number of subcarriers & $512$ \\
    Subcarrier spacing & $15$ kHz \\
    BS transmit power & $43$ dBm \\
    White noise power density & $-174$ dBm/Hz \\
    Intercell distant & $2$ km \\
    Target BER & $10^{-6}$ \\\hline
  \end{tabular} \\
  \label{tblsimparam}
\end{table}

\begin{figure}[h]
  \centering
  \includegraphics[width=3.72in]{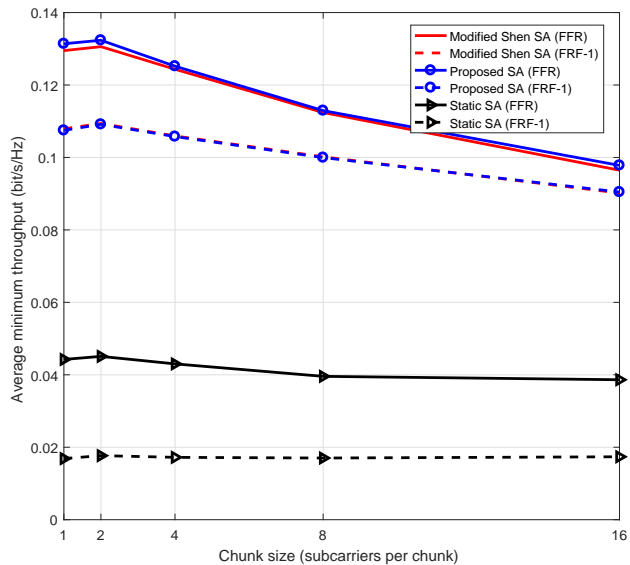}
  \caption{Average minimum throughput of cell-edge users in cell 1
    from various SA schemes at various chunk sizes.}
  \label{fig:multicell_rate}
\end{figure}

In multi-cell setting, an important performance measure is the
throughput of cell-edge users.  In this simulation, the threshold
$\tau$ is set to $0.5R$, and the average minimum throughput of
cell-edge users in cell 1 obtained by various SA schemes at various
chunk sizes are observed.  We modify the single subcarrier-based SA
algorithm proposed by~\cite{journal:shen} by replacing the rate of
individual subcarrier in the algorithm with the average rate over a
chunk.  Results are shown in Fig.~\ref{fig:multicell_rate}.

\begin{figure}[h]
  \centering
  \includegraphics[width=3.72in]{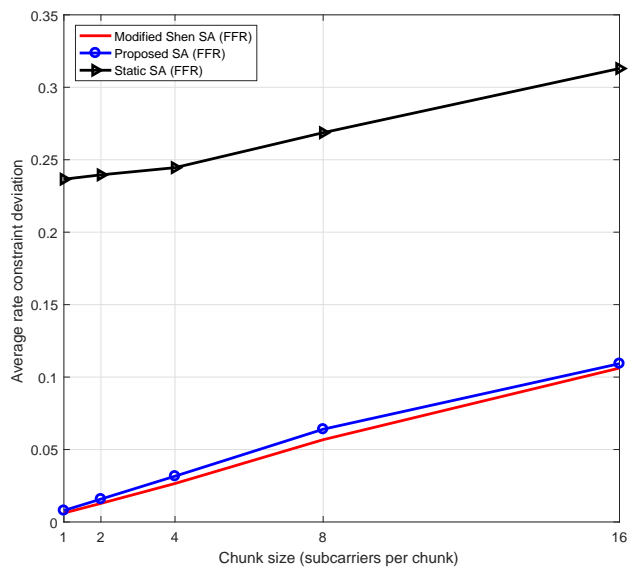}
  \caption{Rate constraint deviation of cell-edge rates from various
    SA schemes at various chunk sizes.}
  \label{fig:multicell_dev}
\end{figure}

In all SA schemes, the throughputs are higher when FFR is used.  The
proposed SA is able to achieve the highest throughput for every chunk
sizes.  The difference in rate is pronounced when compared with static
SA (the rate gain can be as large as 400\%), but is not much when
compared with the SA scheme modified from Shen {\em et al.}'s.  As
chunk size increases, the rate performance of the cell-edge users
decreases.  However, if chunk size is set to 4 or 8, the rate loss is
not significant, but the complexity of SA can be reduced
significantly.  In Fig.~\ref{fig:multicell_dev}, average rate
constraint deviation of cell-edge users is also plotted.  As expected,
the deviation increases as chunk size increases.  From the results,
static SA has the highest deviation since the subcarriers are not
adaptively assigned.  The proposed SA and modified Shen {\em et al.}'s
SA give much lower deviation and thus, fairer.

\section{Conclusions}

In this work, we have proposed a chunk-based SA with PA and
proportional-rate constraints for downlink OFDMA.  Numerical results
show that our proposed SA is able to obtain higher average minimum
user's achievable rate than existing schemes for both
single-subcarrier-based and chunk-based assignment.  In low SNR
regime, user rate is more sensitive to PA and is at its highest with
the proposed PA.  However, in high SNR regime, user rate is more
sensitive to SA and the rate obtained from different PA's does not
differ much as expected.  With single-subcarrier-based assignment,
uniform PA is sufficient to satisfy the rate constraints. However,
with larger subcarrier chunks, proportional rates are more difficult
to maintain with only SA.  Thus, the proposed PA is required to reduce
the effects of low-gain subcarriers within the chunk.

In a multi-cell scenario, a key parameter affecting system performance
is the cell-centre radius.  Determining the proper cell-centre radius
is important, and must be done prior to SA.  Results show that the
proposed SA outperforms existing methods and that static SA has the
worst performance among all SA's.  In addition, implementing FFR does
improve the cell-edge user performance; however, as chunk size grows
larger, the performance gain will be less noticeable.

The proposed scheme relies on the channel information, which can be
accurately estimated at the mobiles and fed back to the base
station. If estimation or feedback errors are significant, the
performance will suffer.  Adaptive resource allocation considered in
this work is appropriate when channel is not very dynamic.  Otherwise,
the system will need to re-compute SA and PA more often.  Effect of
estimation or feedback error or channel's fade rate on the performance
can be analysed in future work.
  
\section*{Acknowledgements}

This work was supported by Kasetsart University Research and
Development Institute (KURDI) under the FY2016 Kasetsart University
research grant and a joint funding from the Thailand Commission on
Higher Education, Thailand Research Fund, and Kasetsart University
under grant number MRG5580236.

\bibliographystyle{IEEEtran}
\bibliography{IEEEabrv,reference}

\end{document}